\documentclass[aps,prl,twocolumn,floatfix,superscriptaddress,showpacs,amsfonts,amssymb,amsmath,preprintnumbers]{revtex4}
\usepackage{bm}
\usepackage{epsfig}
\usepackage{graphicx}
\usepackage{color}
\usepackage{subfigure}

\newcommand{\fig}[1]{Fig.~\ref{#1}}
\newcommand{\Fig}[1]{Fig.~\ref{#1}}
 
\newcommand{\tcl}{t_{\rm cl}}
\newcommand{\tej}{t_{\rm ej}}
\newcommand{\Eeff}{E_{\rm eff}}
\newcommand{\normEeff}{\tilde E_{\rm eff}}
\renewcommand{\d}{\partial}
\newcommand{\beq}{\begin{equation}}
\newcommand{\eeq}{\end{equation}}

\begin{document}

\title{Magnetic reconnection and stochastic plasmoid chains in high-Lundquist-number plasmas}
\author{N.\ F.\ Loureiro}
\affiliation{Associa\c{c}\~ao EURATOM/IST, Instituto de Plasmas e Fus\~ao Nuclear --- 
Laborat\'orio Associado,\\
Instituto Superior T\'ecnico, 
Universidade T\'ecnica de Lisboa,
1049-001 Lisboa, Portugal}
\author{R.\ Samtaney}
%Division of Physical Sciences and Engineering 
%\& Division of Mathematical and Computational Sciences and Engineering, 
\affiliation{King Abdullah University of Science and Technology, Thuwal 23955, Saudi Arabia}
\author{A.\ A.\ Schekochihin}
\affiliation{Rudolf Peierls Centre for Theoretical Physics, University of Oxford, Oxford OX1 3NP, UK}
\author{D.\ A.\ Uzdensky}
\affiliation{Center for Integrated Plasma Studies, %Physics Department, 
University of Colorado, Boulder CO 80309, USA}

\date{\today}

\begin{abstract}
A numerical study of magnetic reconnection in the 
large-Lundquist-number ($S$), plasmoid-dominated regime is carried out 
for $S$ up to $10^7$. The theoretical 
model of Uzdensky {\it et al.} [Phys. Rev. Lett. {\bf 105}, 235002 (2010)] is 
confirmed and partially amended. The normalized reconnection 
rate is $\normEeff\sim 0.02$ independently of $S$ for $S\gg10^4$. 
The plasmoid %system is characterized by 
flux ($\Psi$) and half-width ($w_x$) distribution functions
scale as $f(\Psi)\sim \Psi^{-2}$ and $f(w_x)\sim w_x^{-2}$. 
The joint distribution of $\Psi$ and $w_x$ shows that plasmoids 
populate a triangular region $w_x\gtrsim\Psi/B_0$, where $B_0$ is the 
reconnecting field. It is argued that this feature is due to 
plasmoid coalescence. 
Macroscopic ``monster'' plasmoids with $w_x\sim 10$\% of the system size  
are shown to emerge in just a few Alfv\'en times, independently of $S$, 
suggesting that large disruptive events are an inevitable feature of 
large-$S$ reconnection. 

\end{abstract}

%NFL: I think only 4 pacs are allowed, 52.65.Kj}

\maketitle

%***********************************************************************************

\paragraph{Introduction.} 

It has become clear in recent years that resistive magnetic reconnection at asymptotically
high Lundquist numbers ($S$) is a temporally and spatially irregular process, dominated by multiple
plasmoids generated in unstable current 
sheets~\cite{Loureiro_07,Lapenta_08,Daughton_09,Loureiro_09,Samtaney_09,Bhatta_09,Cassak_09,Huang_10}. 
The reconnection rate in this regime is independent of $S$ provided 
$S>S_c$ \cite{Bhatta_09,Huang_10}, 
where $S_c\sim10^4$ \cite{Biskamp_86,Loureiro_05} is the plasmoid 
instability \cite{Loureiro_07} threshold.
Thus, the classic Sweet-Parker (SP) theory \cite{Sweet_58,Parker_57} 
is no longer sufficient even for resistive MHD reconnection and a 
new physical paradigm is needed.

Such a theory was recently attempted by \cite{Uzdensky_10} 
(henceforth ULS).
The physical picture on which it is based is that, as the plasmoid 
instability \cite{Loureiro_07}
proceeds into its nonlinear stage, inter-plasmoid current sheets form, which 
are then subject to the same instability. The result is 
a multiscale plasmoid chain originally envisioned by \cite{Shibata_01}. 
ULS assume that (i) the current sheets connecting the plasmoids in this chain
are typically just marginal with respect to the plasmoid instability and 
so their length is $\sim L_c=(\eta/V_A)S_c$, where $\eta$ is the magnetic diffusivity 
and $V_A$ the Alfv\'en speed based on the upstream magnetic field $B_0$;
(ii) the reconnecting field is equal to the upstream field $B_0$ 
for all interplasmoid layers and so 
outflows into all plasmoids are Alfv\'enic with the same speed $V_A$; 
and (iii) smaller plasmoids do not have time to saturate before they are ejected 
into larger ones (and are promptly merged with them). 
ULS then show that (i) the effective reconnection rate 
is $\normEeff = c\Eeff/B_0V_A\sim S_c^{-1/2}\sim 0.01$; 
(ii) the plasmoid flux ($\Psi$) and cross-sheet half-width ($w_x$)
distribution functions are $f(\Psi)\sim \normEeff B_0 L \Psi^{-2}$ and
$f(w_x)\sim \normEeff L w_x^{-2}$ (the power laws are the same because it is 
argued that $\Psi \sim w_x B_0$); 
and (iii) anomalously large ``monster'' plasmoids occasionally occur, 
with sizes $\sim S_c^{-1/4}L \sim 0.1L$, where $L$ is the system size. 
Note that diagnosing the plasmoid chain in terms of the flux and half-width 
distributions is a natural statistical description for such an object 
\cite{Fermo_10,Uzdensky_10,Nishizuka_09}.
Note also that the prediction of monster plasmoids is potentially an important one 
in light of the evidence of violent
abrupt events associated with reconnection sites (e.g., solar flares \cite{Lin_05} 
or sawtooth \cite{Park_06}).

In this Letter, we present a numerical study
of resistive MHD reconnection at the highest currently achievable Lundquist numbers.
Our results confirm the basic predictions of the ULS theory, but also reveal 
that the picture is more complex than originally envisioned. 

\begin{figure*}[t!]
\hskip 0.7cm
 \includegraphics[angle=-90,width=16.1cm]{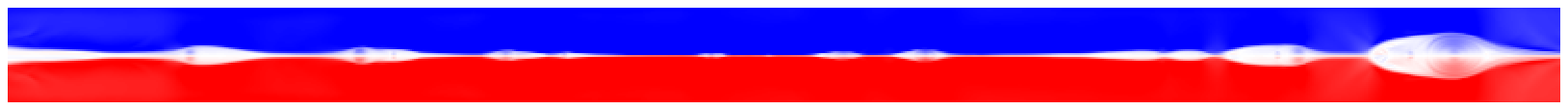}
\vskip -0.4cm
\includegraphics[width=18cm]{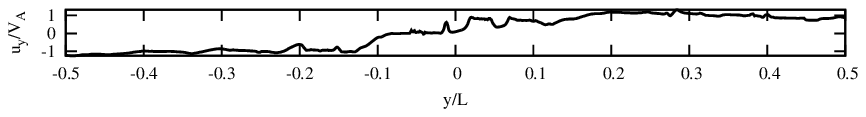}
\vskip-0.25cm
\caption{\label{fig:snaps} Top panel: the plasmoid chain for the run with $S=10^6$ ($16384^2$ 
grid points). 
Only a fraction of the $x$-domain is shown, 
%In the $x$-direction (vertical) only a fraction of the domain is shown: 
$-0.03\le x/L\le0.03$.
The color scale (blue to red) represents $B_y\in[-1,1]$. 
Note that the assumption of ULS \cite{Uzdensky_10} that the reconnecting 
field is equal to the upstream field $B_0$ all the way to the 
thinnest of the current sheets appears to hold true. 
Bottom panel: outflow velocity $u_y(x=0,y)$. The outflows into most plasmoids 
are approximately Alfv\'enic.}
\vskip-0.25cm
\end{figure*}

\paragraph{Numerical setup.}
We use the same numerical scheme as \cite{Samtaney_09}
to solve the standard set of compressible visco-resistive MHD equations in a 2D box 
$[-L_x,L_x]\times[-L_y,L_y]$. 
Our setup is designed so that a statistical steady state can be reached. Namely, 
the density, pressure and the incoming magnetic field are imposed at the upstream 
boundaries ($x=\pm L_x$): $\rho=1$, $P=3$ and 
$B_y=B_0\left\{1+\cos[(\pi y/2 L+\epsilon)^2]\right\}/2$, 
where the code units are based on $V_A=B_0/\sqrt{4\pi\rho}=1$ and $L=1$.
The small perturbation $\epsilon=0.06L_y/L$ is necessary to break the $y$-symmetry of the 
numerical set up and thus prevent the artifical lingering of plasmoids at the center of the sheet.
Solenoidality is used to fix $\d B_x/\d x = - \d B_y/\d y$ at the upstream boundary. 
For the velocity at this boundary, we set $\d u_y/\d x=0$,
whereas $u_x(x=\pm L_x)$ is obtained from the frozen-flux condition 
(the box is wide enough that the resistive term is negligible at $x=\pm L_x$). 
Free-outflow boundary conditions are imposed at the downstream ($y=\pm L_y$) boundaries.
The initial condition is designed to mimic qualitatively a SP-like current sheet. 
This is not, however, a steady-state solution 
of the resistive MHD equations, so there is no need to add a perturbation to the 
initial configuration in order to trigger the plasmoid instability. 
The instability threshold for this setup is found to be $S_c\approx 1.2\times10^4$.

We perform a Lundquist-number scan in the range $300\le S \le 10^7$.
In all cases, the viscosity $\nu=\eta$. 
Most of our runs are done in a ``semi-global'' setup with $L_x=0.3L$ and $L_y=0.5L$;
the exceptions, for lack of sufficient computational resources, 
are the runs with $S=3\times10^6$ ($L_x=0.15L,~L_y=0.25L$) and 
$S=10^7$ ($L_x=0.01L,~L_y=0.02L$) \footnote{Even in this case, the box is still much 
wider (in $x$) than the thickness of the SP layer, $\delta_{\rm SP}/L\sim S^{-1/2}\sim 3\times10^{-4}$, 
and much longer (in $y$) than $L_c/L \sim S_c/S \sim 10^{-3}$.}. 
%These are of limited use in terms of measuring the global plasmoid 
%distribution functions, but are valid for determining the reconnection rate at 
%the highest values of $S$ available so far. 
The numerical resolution depends on $S$, ranging up to 
$16384^2$ for $S=10^6,~3\times 10^6$ and $4096\times8192$ for $S=10^7$
(for which the box is smaller).

\begin{figure}[b!]
\includegraphics[width=8cm]{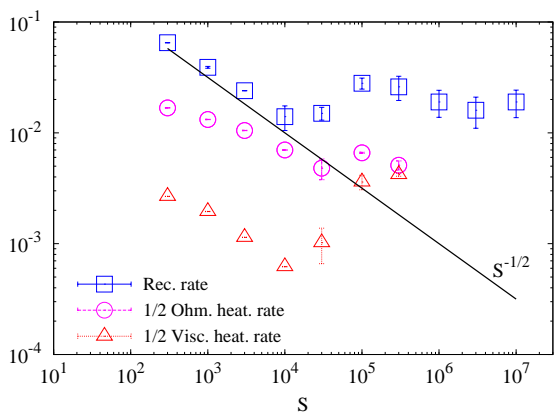}
\vskip-0.25cm
\caption{Reconnection rate $\normEeff$ [Eq.~\eqref{eq:rec_rate}, squares], 
and the (half) rates 
of resistive $Q_\eta$ (circles) and viscous $Q_\nu$ (triangles) heating. 
Simulations with $S\ge10^6$ last for shorter times before they are disrupted by 
monster ejections and so converged mean values for heating rates could not be obtained; 
the reconnection rate, calculated at the inflow boundary, did not have this problem.}
\label{fig:rec_rate}
\end{figure}

\paragraph{Reconnection rate.}
In all our simulations with $S>S_c$, the initial SP-like configuration 
is quickly replaced with a plasmoid-dominated current sheet (\fig{fig:snaps}). 
The system then enters a statistical steady state, with multiple plasmoids constantly 
being formed, coalescing and being ejected through the outflow boundary.
We define the effective global reconnection rate in terms the 
inflow plasma velocity at the upstream wall: 
%$\normEeff = \langle\int_{-L_y}^{L_y}\!dy\, u_x(x=L_x,y)/2L_yV_A\rangle,$
\begin{equation}
\normEeff = \left\langle\frac{1}{2L_yV_A}\int\!dy\, u_x(x=L_x,y)\right\rangle,
\label{eq:rec_rate}
\end{equation} 
where $\langle\dots\rangle$ denotes time average.   
This is plotted in \fig{fig:rec_rate} as a function of $S$. 
A transition is manifest from the SP scaling $\normEeff\sim S^{-1/2}$ for 
$S\lesssim 10^4\sim S_c$ to $\normEeff\approx 0.02 \sim S_c^{-1/2}$, 
consistent with the ULS prediction \cite{Uzdensky_10}
and previous numerical results \cite{Daughton_09,Loureiro_09,Bhatta_09,Cassak_09,Huang_10}. 
Note that this result is now extended to larger values of $S$ than ever before. 
Such an extension is important: 
as shown by ULS, $S\sim S_c^{3/2}\sim10^6$ is the threshold at which an individual 
plasmoid can saturate faster than it is ejected from the global current sheet. 
This would slow down reconnection were it not for plasmoid ejection: 
smaller plasmoids are swallowed by larger ones before they have time to saturate.
It was assumed by ULS that this coalescence process would operate efficiently 
--- the persistence of fast reconnection beyond $S\sim10^6$ demonstrated here 
suggests that this assumption is indeed valid. 

\paragraph{Heating rate.} The normalized resistive and viscous heating rates are  
$Q_\eta = \langle\int\!\!\int\! dxdy\,\eta j_z^2(x,y)/2L_yB_0^2V_A\rangle$, 
and $Q_\nu = \langle\int\!\!\int\! dxdy\,\nu \omega_z^2(x,y)/2L_yV_A^3\rangle$, 
%\begin{equation}
%Q_\eta = \left\langle\frac{1}{2L_yB_0^2V_A}\int\!\!\int\! dxdy\,\eta j^2(x,y)\right\rangle, 
%\label{eq:res}
%\end{equation}
where $j_z$ and $\omega_z$ are the current and vorticity, respectively. 
These rates 
%and the total (kinetic $+$ magnetic) energy outflux through the $y=\pm L_y$ boundaries 
%$P_{\rm out}= 2\langle\int\! dx\left[\rho u^2u_y/2 + 
%cE_zB_x/4\pi\right](x,y=L_y)/\rho V_A^3\rangle$, 
are also plotted in \fig{fig:rec_rate}.
In the fast-reconnection regime, $Q_\eta\approx Q_\nu \approx 0.008$. 
Since the total Poynting flux into the box is (per unit length)
$P_{\rm in} \approx 2\normEeff \approx 0.04$
and the kinetic energy influx is small ($\propto u_x^3 ~ \normEeff^3$), 
the conclusion is that $\sim 40\%$ of the incoming (magnetic) energy 
is dissipated into heat (the rest goes into the reconnected field 
and the kinetic energy of the mass outflows). 

\begin{figure*}[t!]
\includegraphics[width=8cm]{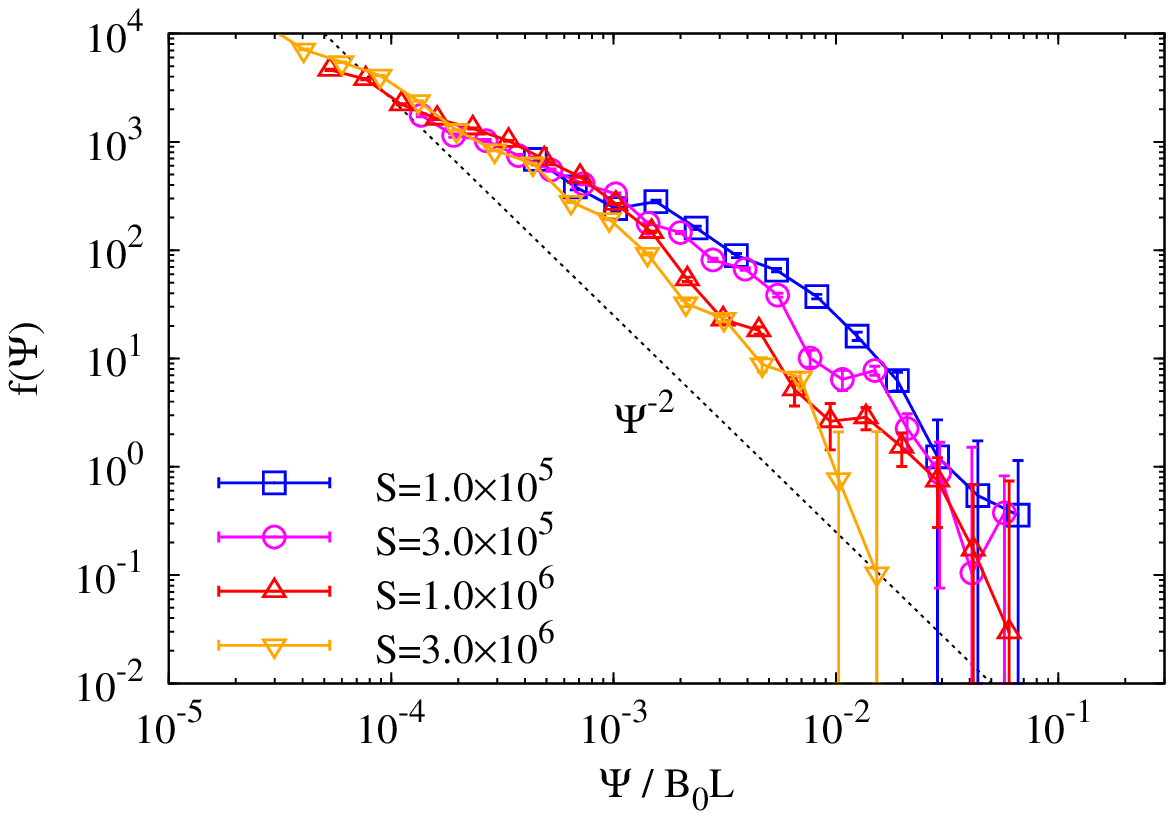}
\includegraphics[width=8cm]{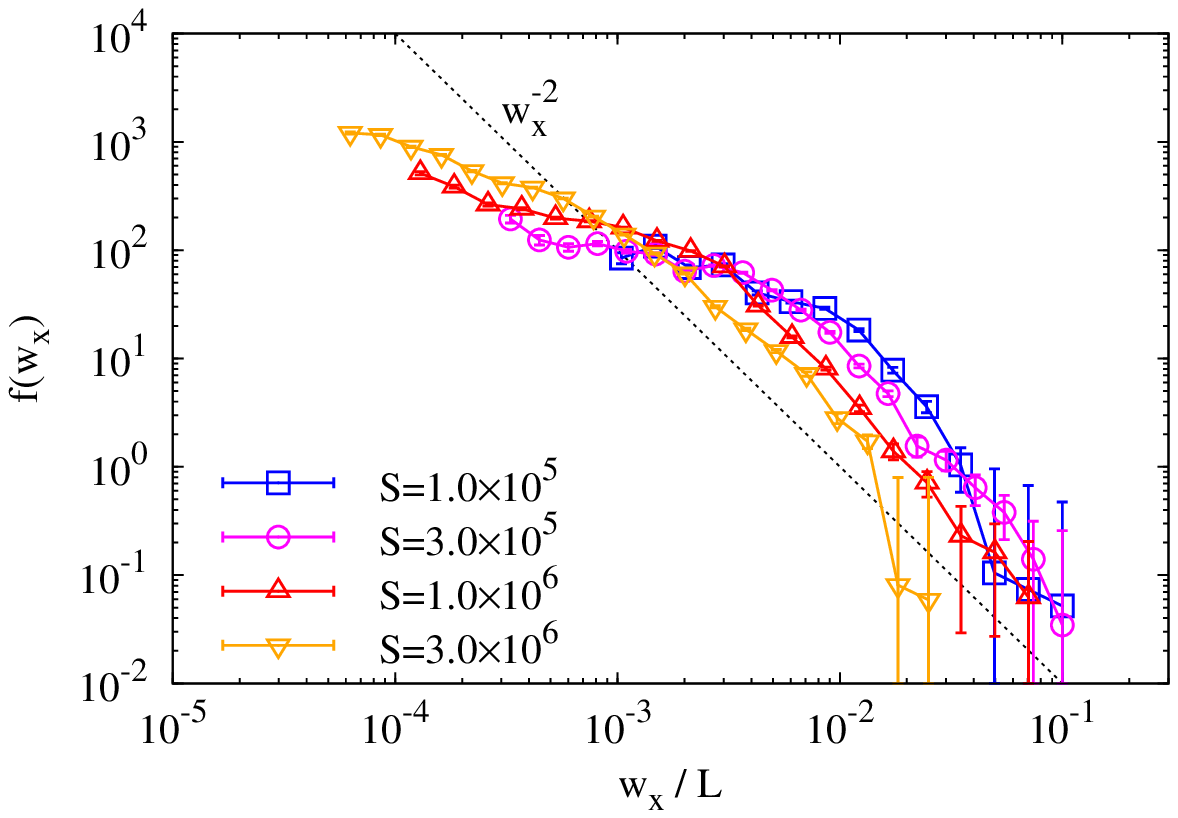}
\vskip-0.5cm
\caption{Plasmoid flux (left) and half-width (right) distribution functions
(note that the run with $S=3\times10^6$ had a shorter box by half and so 
its distributions cut off at smaller flux and width). Dashed
lines show the ULS scalings~\cite{Uzdensky_10}. 
%$f(\Psi) \propto \Psi^{-2}$ and $f(w_x) \propto w_x^{-2}$.
%$f(\Psi) = (S_c/S)\normEeff B_0 L \Psi^{-2}$ and $f(w_x) = (S_c/S)\normEeff L w_x^{-2}$.
%The vertical lines show the knee position $w_{x*}$ estimated in \eqref{eq:knee}.
}
\vskip-0.25cm
\label{fig:dist_funcs}
\end{figure*}

\begin{figure}[b!]
\vskip-0.5cm
\includegraphics[width=8.5cm]{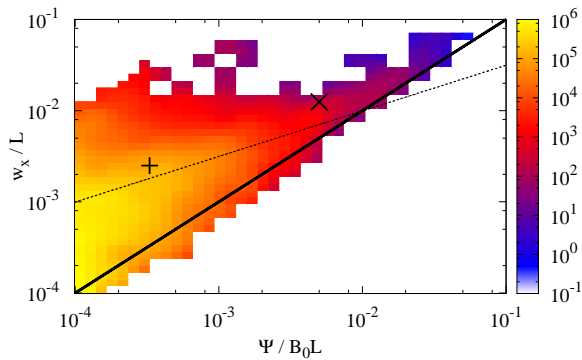}
\vskip-0.5cm
\caption{Joint distribution of plasmoid flux and half-width for $S=10^6$. 
The solid diagonal line shows the ULS plasmoids $w_x=\Psi/B_0$; the dotted line 
is the condition \eqref{eq:max}. The predator and prey plasmoids shown in 
\fig{fig:contour_zoom} are marked by $\times$ and $+$, respectively.}
\label{fig:triangle}
\end{figure}

\paragraph{Plasmoid distribution.}
The plasmoid population is naturally characterized by the distribution 
of fluxes ($\Psi$) and half-widths ($w_x$) of individual plasmoids \cite{Fermo_10,Uzdensky_10}. 
The distribution functions $f(\Psi)$ and $f(w_x)$ are plotted in \fig{fig:dist_funcs}. 
The $\Psi^{-2}$ and $w_x^{-2}$ scalings predicted by ULS do hold, although 
the distributions flatten for $\Psi/B_0L$ and $w_x/L$ below certain values 
that decrease at larger $S$. 
%The origin of this new feature becomes clear if we consider 

A more detailed diagnostic is 
the joint distribution function $f(\Psi,w_x)$, which is shown in \fig{fig:triangle}
and reveals a new feature: 
ULS argued that the plasmoid half-width and flux should be related by 
$w_x\sim \Psi/B_0$; in fact, there is a significant 
off-diagonal plasmoid population with $w_x>\Psi/B_0$ (cf.~\cite{Fermo_10}). 
The presence of these plasmoids in the measured distribution 
can be explained as follows. The ULS argument assumed effectively 
that once a smaller plasmoid is ejected into a larger one, 
it is immediately and completely absorbed by (i.e., coalesces with) 
the latter and so falls out of the distribution. However, in reality, the 
coalescence between two plasmoids is not instantaneous (cf.\ \cite{Fermo_10})
--- and so at any given time, there are many plasmoids for which 
coalescence has started at some earlier time and that 
are in an advanced stage of being digested by a bigger plasmoid. 
Thus, a typical plasmoid's life consists of two distinct phases: the ULS growth 
phase (while the plasmoid moves through its host current sheet) and the subsequent 
phase of digestion by a bigger plasmoid --- this will have an effect on the plasmoid 
distribution. 

\begin{figure}[b!]
\includegraphics[width=8cm]{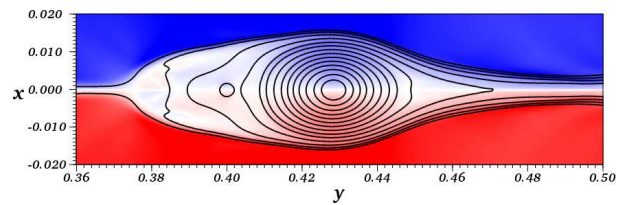}
\caption{Example of coalescing plasmoids: zoom on the rightmost part of~\fig{fig:snaps}.  
Same color scheme is used. Lines of constant magnetic flux are also shown.}
\label{fig:contour_zoom}
\end{figure}

We envision the coalescence  
as a gradual stripping of the outer layers of the smaller plasmoid so  
the magnetic field in a semidigested plasmoid is
$B\sim B_0 w_x/w_{x0}$, where $w_{x0}$ is the plasmoid's half-width at 
the beginning of the coalescence \footnote{
When a typical plasmoid is ejected from its host layer into a bigger plasmoid, it 
is long and thin:  $w_x \sim w_y \normEeff$ \cite{Uzdensky_10}. However, immediately 
upon ejection, it is squashed against the bigger plasmoid and becomes 
circularized ($w_x \sim w_y$), while preserving its flux and area. Thus, just 
before it starts coalescing with the bigger plasmoid, its width jumps up by a 
factor of $\normEeff^{-1/2} \sim 10$, and its magnetic field becomes 
$B_0 \normEeff^{1/2}$ to preserve flux. This 
immediately moves the plasmoid vertically upwards by a factor $\normEeff^{-1/2}$ 
from the ULS diagonal. Our arguments can be modified 
to account for this effect --- the result is to raise the threshold 
\eqref{eq:max} upwards by a factor of order unity.}
%The plasmoids that are not circularized before they are ejected from their sheet 
%will be squashed against 
%the bigger plasmoid before coalescence starts. Initially, the
%ratio between their $x$ and $y$ widths is $\sim\normEeff$ \cite{Uzdensky_10}, 
%so after squashing, their width is $w_{x0}\normEeff^{-1/2}$ and 
%the magnetic field in them is (from flux conservation) $B_0\normEeff^{1/2}$. 
%Our arguments below can be modified accordingly, but this merely
%introduces some relatively insignificant prefactors. Note 
%that the squashing moves the plasmoids vertically upwards by a factor of 
%$\normEeff^{-1/2}$ from the ULS diagonal.}. 
Its flux is, therefore, $\Psi\sim B_0 w_x^2/w_{x0}$.
%\begin{equation}
%\Psi\sim B_0 w_x^2/w_{x0}, \quad w_x<w_{x0}, 
%\label{eq:flux}
%\end{equation}
Since $w_x<w_{x0}$, these plasmoids are off-diagonal: $w_x>\Psi/B_0$. \Fig{fig:contour_zoom} 
illustrates the swallowing of a smaller plasmoid by a larger one; 
as shown in \fig{fig:triangle}, the latter is relatively close to the diagonal, 
while the former is strongly off-diagonal. 

Let us estimate the widths of the off-diagonal plasmoids. 
These have to be relatively small because 
larger plasmoids take a longer time to be digested and if that time exceeds the 
typical time $\tau_A\sim L/V_A$ for both predator and prey plasmoids 
to be ejected from the global sheet, then the effect on the measured distribution 
is small. The characteristic coalescence time is $\tcl\sim \Psi_0/cE \sim B_0w_{x0}/cE$, 
where $\Psi_0$ is the initial flux, 
$cE\sim V_AB_0\max(S_c^{-1/2},S_w^{-1/2})$ is the reconnection rate, 
$S_w\sim V_A w_{x0}/\eta$ is the Lundquist number associated with 
the (vertical) current sheet that forms between two coalescing plasmoids 
and we are taking into account that reconnection rate is 
independent of~$S_w$ for $S_w>S_c$, or $w_{x0}>L_c$ (length of the 
longest possible plasmoid-stable layer \cite{Uzdensky_10}; cf.\ \cite{Barta_11}). 
Therefore, $\tcl/\tau_A \sim S_c^{1/2} (w_{x0}/L)\min(1,\sqrt{w_{x0}/L_c}) \lesssim 1$, 
or $w_{x0}\lesssim L\max(S_c^{-1/2},S^{-1/3})$, is the condition for semidigested plasmoids 
to contribute to the off-diagonal part of the distribution. %Using \eqref{eq:flux}, 
Since $\Psi\sim B_0 w_x^2/w_{x0}$, this translates into
\begin{equation}
w_x/L \lesssim (\Psi/B_0L)^{1/2}\max(S_c^{-1/4},S^{-1/6}). 
\label{eq:max}
\end{equation}
This indeed appears to capture the maximum of $f(\Psi,w_x)$ 
rather well (see \fig{fig:triangle} for $S=10^6$; similarly good agreement 
was found for other values of $S$). 
Since \eqref{eq:max} must be consistent with $w_x>\Psi/B_0$,
the off-diagonal plasmoids only matter if 
$w_x/L \lesssim \max (S_c^{-1/2},S^{-1/3})$. 
%In \fig{fig:dist_funcs}, this
%is compared with the location of the knee in $f(w_x)$. 
Note that the coalescence rate becomes 
independent of $\eta$ for $S\gtrsim S_c^{3/2}\sim10^6$. 

\begin{figure}[t!]
\includegraphics[width=8cm]{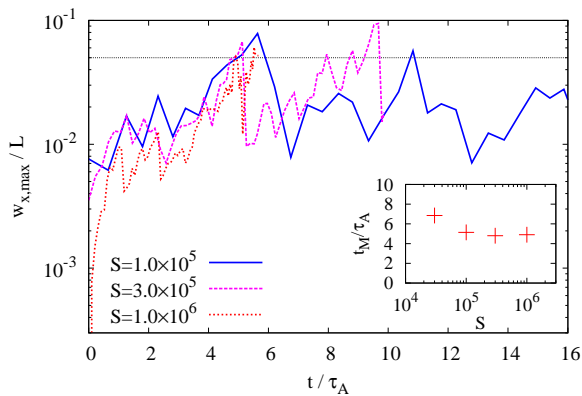}
\vskip-0.5cm
\caption{The half-width of the largest plasmoid in the box vs.\ time. 
Time is measured from the start of steady-reconnection-rate phase.
The dotted line is the monster threshold $0.05L$. 
Inset: time $t_M$ to reach the monster threshold vs.~$S$.}
\vskip-0.25cm
\label{fig:monsters}
\end{figure}

\paragraph{Monster plasmoids.} The following argument follows ULS \cite{Uzdensky_10}, 
in a somewhat expanded and amended form. Because the flows carrying both flux and 
embedded plasmoids out of the current sheet are roughly linear, $u_y\sim V_A y/L$
(see \fig{fig:snaps}), the ejection time for a plasmoid born at some 
location $y_0$ in the sheet is $\tej=\int_{y_0}^L dy/u_y \sim \tau_A \ln (L/y_0)$
(this is true both for the global sheet and also for any local one, in which 
case $L$ would be the typical length for the latter). 
Therefore, plasmoids born near the center of the sheet 
remain in the game logarithmically longer than others. 
While this only leads to a logarithmic correction for their 
flux, $\Psi\sim \normEeff V_AB_0\tej\sim \normEeff B_0L\ln L/y_0$, 
the enhancement of their area is much greater. 
A plasmoid grows in area by absorbing all the plasma and 
smaller plasmoids from roughly up to the midpoint of the layer that connects 
it to its neighbor of a similar size. 
One can then see that the plasmoid area $A$ grows according to
\begin{equation} 
\frac{dA}{dt} \sim \Delta y(t) \frac{d}{dt}\frac{\Psi}{B_0} 
\sim \Delta y(0) e^{t/\tau_A} \normEeff V_A,
\label{eq:area}
\end{equation}
where $\Delta y(t)$ is the (exponentially stretched) half-distance 
to the neighboring plasmoid. 
Integrating \eqref{eq:area} up to $t=\tej$ gives 
$A(\tej)\sim \normEeff \Delta y(0) L(L/y_0-1)$.
 
If the plasmoid was born away from the center 
of the sheet, $y_0\sim L$, then $A\sim \normEeff \Delta y(0) L$
and so $w_x\sim A/w_y\sim\normEeff L\sim 0.01 L$. We have estimated the $y$-extent 
of the plasmoid as $w_y\sim\Delta y(0)$, which does not change as long as $w_x<w_y$. 
In contrast, for centrally born plasmoids, $y_0\ll L$, we have
$w_x\sim \normEeff L^2/y_0$ at ejection, 
provided $w_x< w_y\sim\Delta y(0)$. 
If the latter condition is not satisfied, i.e., if $y_0 < \normEeff L^2/\Delta y(0)$, 
the plasmoid will be circularized as soon as $w_x\sim w_y$ (which will happen 
before ejection) and so its half-width at ejection will be 
$w_x\sim\normEeff^{1/2}L (\Delta y(0)/y_0)^{1/2}$. 
Since $\Delta y(0)\le y_0$, the maximum half-width achievable is 
$w_{x,\rm max}\sim \normEeff^{1/2}L\sim 0.1L$. 
This is a nearly macroscopic size --- the plasmoids that reach 
it were dubbed {\em monster plasmoids} by ULS. 
Only those plasmoids stand a chance of achieving monster status that 
are born at $y_0 < \normEeff^{1/2} L\sim 0.1L$. 
This must be consistent with $y_0\gtrsim L_c$ 
(shorter sheets are stable), which implies that monsters will only appear 
if $S\gtrsim S_c^{5/4}\sim10^5$. 

\fig{fig:monsters} shows the half-width of the largest 
plasmoid, $w_{x,{\rm max}}$ in the simulation box vs.\ time. 
Exponential growth to the monster size is manifest --- this 
is defined here, somewhat arbitrarily, as $w_{x,{\rm max}}=0.05L$ 
($0.1L$ is never actually reached in our simulations, but it is, 
of course, no more than an order-of-magnitude estimate; also 
our simulation domain is smaller than the system size, $L_y<L$). 
This size is usually achieved by just one plasmoid at a time, 
just before it is ejected, whereupon $w_{x,{\rm max}}$ dips, 
then recovers as a new monster emerges, and so on. 
The time $t_M$ for a plasmoid system to produce 
and grow a monster can be estimated simply as the ejection 
time for a plasmoid born in the relevant central part of the sheet: 
$t_M\sim\tej\sim\tau_A\ln\normEeff^{-1/2}$, which amounts to a few Alfv\'en times,
independent of $S$ --- this is borne out by the numerical results
(\fig{fig:monsters}, inset). 
For monster plasmoids, $\Psi < B_0 w_x$ (like for the coalescing ones),
so they occupy the top right corner of the $(\Psi,w_x)$ plane 
in \fig{fig:triangle} (note that the large plasmoid in \fig{fig:contour_zoom}
is a monster in the making). The probability of finding a monster 
(defined by $w_x>0.05L$) hovers between 1\% and 3\%.  

\paragraph{Conclusions.} 
We have found that resistive MHD reconnection 
is fast, its rate $cE/B_0 V_A=\normEeff\sim 0.02$, independently of $S$.
While a similar conclusion has been reported before 
\cite{Daughton_09,Loureiro_09,Bhatta_09,Cassak_09,Huang_10}, 
our study is the first to probe Lundquist numbers significantly exceeding 
the critical threshold of $10^6$ \cite{Uzdensky_10} in order to show
that plasmoid saturation does not shut down fast reconnection  
in the high-Lundquist-number, plasmoid-mediated regime. It also confirms 
that reconnection occurs via a multiscale plasmoid chain 
\cite{Shibata_01,Fermo_10,Uzdensky_10,Barta_11}, 
characterized by local Alfv\'enic outflows and many coalescing plasmoids. 

Statistics of this ``plasmoid turbulence'' are measured 
for the first time in terms of the flux-width joint distribution --- a natural 
choice both from the theoretical \cite{Fermo_10,Uzdensky_10} and 
observational \cite{Nishizuka_09} perspective. 
The ULS scalings $w_x\sim\Psi/B_0$, $f(\Psi)\sim \Psi^{-2}$, $f(w_x)\sim w_x^{-2}$ 
are corroborated, but we also find a substantial ``off-diagonal'' ($w_x>\Psi/B_0$) 
plasmoid population 
for $w_x \lesssim \normEeff L$. The excess of plasmoids of relatively large size 
and small flux is explained by considering the coalescence 
between plasmoids. Thus, the full picture of the plasmoid ``turbulence'' involves 
not just multiple reconnection sites along the global layer, but also 
many transverse layers between coalescing plasmoids (these layers can themselves 
break up into plasmoid chains \cite{Barta_11}). 

Another large-size low-flux subspecies is the ``monster'' plasmoids, also 
theoretically anticipated by \cite{Uzdensky_10}. They are born in the middle tenth  
of the global layer and grow to nearly macroscopic size in just a few Alfv\'en times, 
independently of the Lundquist number. 
This inevitable and relatively frequent nature of what can be 
very violent and disruptive events (ejection of a monster from the 
global layer) is reminiscent of the observed bursty character 
of plasmoid ejections  in solar flares \cite{Lin_05,Karlicky_10} and perhaps 
also of the sawtooth crash in tokamaks \cite{Park_06}. 
%For example, it resonates strongly with the experimentally verified
%independence of the sawtooth's crash time on the Lundquist number~\cite{Hastie_97},
%though this observation is strictly of a speculative nature at this stage.

These results show that even 2D MHD resistive reconnection contains a wealth 
of strongly nonlinear, stochastic behavior --- a type of MHD turbulence 
that is only now starting to be studied quantitatively. It is encouraging 
that the simple phenomenology of the ULS model \cite{Uzdensky_10} appears 
to capture some of the essential properties of such systems, 
but it is also now clear that the full picture will require a deeper 
and more quantitative understanding of plasmoid coalescence and of the 
extreme events such as the emergence of monsters. 
Finally, many further complications will have to be taken into account 
before idealized models can truly describe the real-world reconnection 
in its full splendor: e.g., 
kinetic physics \cite{Daughton_09,Daughton_11,Karimabadi_11}, 
background turbulence \cite{Lazarian_99,Loureiro_09}, 
3D effects \cite{Lapenta_11,Daughton_11}).

\paragraph{Acknowledgments.}
This work was supported by 
Funda\c{c}\~ao para a Ci\^{e}ncia e Tecnologia and by the European Communities
under the contract of Association between EURATOM and IST
(NFL), STFC (AAS), and by the Leverhulme Network for Magnetized Plasma Turbulence.
The views expressed herein do not necessarily reflect those of the 
European Commission. 
Simulations were carried out at HPC-FF (Juelich), Jugene (Juelich; PRACE grant PRA024), 
Ranger (NCSA) and IBM Blue Gene Shaheen (KAUST).

%-----------------------------------------------------------------------------------

\bibliography{nlplasmoidsims}

\end{document}